\documentclass[manuscript]{acmart}

\setcopyright{rightsretained}

\acmJournal{TOIT} 

\usepackage{paralist}
\usepackage{xargs} 
\usepackage[utf8]{inputenc}


\begin{document}

\title{Governance of Autonomous Agents on the Web: Challenges and Opportunities}


\author[Kampik]{Timotheus Kampik}
\orcid{0000-0002-6458-2252}
\affiliation{%
  \institution{Umeå University}
  \city{Umeå}
  \country{Sweden}}
\email{tkampik@cs.umu.se}

\author[Mansour]{Adnane Mansour}
\affiliation{%
  \institution{Mines Saint-\'Etienne}
  \country{France}}
\email{adnane.mansour@emse.fr}

\author[Boissier]{Olivier Boissier}
\affiliation{%
  \institution{Mines Saint-\'Etienne}
  \country{France}}
\email{olivier.boissier@emse.fr}

\author[Kirrane]{Sabrina Kirrane}
\orcid{0000-0002-6955-7718}
\affiliation{%
  \institution{Wirtschaftsuniversität Wien}
  \city{Vienna}
  \country{Austria}}
\email{sabrina.kirrane@wu.ac.at}

\author[Padget]{Julian Padget}
\orcid{0000-0003-1314-2094}
\affiliation{%
  \institution{University of Bath}
  \city{Bath}
  \country{UK}}
\email{j.a.padget@bath.ac.uk}

\author[Payne]{Terry R. Payne}
\affiliation{%
  \institution{University of Liverpool}
  \city{Liverpool}
  \country{UK}}
\email{t.r.payne@liverpool.ac.uk}

\author[Singh]{Munindar P.~Singh}
\orcid{0000-0003-3599-3893}
\affiliation{%
  \institution{North Carolina State University}
  \city{Raleigh}
  \state{North Carolina}
  \country{USA}
  \postcode{27695}
}
\email{mpsingh@ncsu.edu}

\author[Tamma]{Valentina Tamma}
\affiliation{%
  \institution{University of Liverpool}
  \city{Liverpool}
  \country{UK}}
\email{v.tamma@liverpool.ac.uk}

\author[Zimmermann]{Antoine Zimmermann}
\affiliation{%
  \institution{Mines Saint-\'Etienne}
  \country{France}}
\email{antoine.zimmermann@emse.fr}

\begin{abstract}
The study of autonomous agents has a long tradition in the Multiagent System and the Semantic Web communities, with applications ranging from automating business processes to personal assistants. More recently, the Web of Things (WoT), which is an extension of the Internet of Things (IoT) with metadata expressed in Web standards, and its community provide further motivation for pushing the autonomous agents research agenda forward. Although representing and reasoning about norms, policies and preferences is crucial to ensuring that autonomous agents act in a manner that satisfies stakeholder requirements, normative concepts, policies and preferences have yet to be considered as first-class abstractions in Web-based multiagent systems. Towards this end, this paper motivates the need for alignment and joint research across the Multiagent Systems, Semantic Web, and WoT communities, introduces a conceptual framework for governance of autonomous agents on the Web, and identifies several research challenges and opportunities.
\end{abstract}

\ccsdesc[500]{Computing methodologies~Multiagent Systems}
\ccsdesc[500]{Computing methodologies~Intelligent agents}
\ccsdesc[500]{Information systems~web services}

\keywords{autonomous agents, norms, policies, preferences, governance}

\maketitle


\section{Introduction}
Over the last two decades, the Web has evolved extensively in response to a variety of different requirements. From originally providing a distributed information dissemination architecture, it has encompassed support for publication, discovery, consumption and aggregation of information, knowledge, and services, thereby interconnecting the digital, social and physical worlds. 
The Web's ubiquity, as well as the simplicity of its underlying communication protocols has resulted in it becoming the de facto standard for communication between services, and more recently, connected things.
With the rise of the Internet of Things (IoT), 
the combination of the Web of Things (WoT) (as an extension of the IoT with metadata expressed in Web standards), traditional web services, and a knowledge dissemination infrastructure that is both machine navigable and machine understandable, has facilitated a new generation of applications that utilise the Web. 

\citet{berners2001semantic} outline how autonomous agents could comprehend and exploit this machine-readable knowledge to achieve a variety of tasks. Thus, the notion of \emph{autonomy} provides a framework whereby individual agents (e.g., those representing or controlling services, things, or applications) may plan, collaborate, and cooperate to achieve complex but disparate goals. Such multiagent systems avoid centralised control, which is the bane of business process management~\cite{SOC-05}. By seeking mutually beneficial interactions, agents of heterogeneous construction (potentially originating from different developers) can evolve a mutually supportive economy across the Web, performing a multitude of tasks for Web users.
However, to achieve this notion of collaborative agents that use the Web infrastructure, it is crucial to consider a \emph{governance} perspective, which defines how agents should act in a given situation (also considering the consequences of their potential actions) and defines how frameworks that govern communities of agents should be designed, interoperate, and evolve.
This perspective is of particular importance for the Web, where usage may cross social contexts and jurisdictions, and where no centralised control over the different agents is possible.
Indeed, the need for intelligent system governance is, at the time of writing, a focus point of legislative and regulatory efforts; e.g., by the European Commission~\cite{aihleg}.

Therefore, what is needed is a new governance framework supported by a review of the related literature on the use of norms, policies, and preferences for autonomous normative agents, as well as contextualising these with respect to the notion of the Web (of Things).
Towards this end, this paper makes three main contributions. 
Firstly, it motivates the need for norms, policies and preferences for autonomous agents on the Web by means of a simple motivating scenario.
Secondly, it proposes a new governance conceptual framework and gives an overview of the state of the art on norms, policies, and preferences for autonomous normative agents (restricted to those efforts that provide the theoretical background to our proposed framework).
Finally, it identifies several challenges and opportunities, for the MAS, Semantic Web and WoT communities, underlining the need for better integration and joint research across the different communities.
Each challenge is motivated by a concise review of the state of the art, followed by several opportunities for future investigation. For each of the identified challenges, we discuss its maturity in terms of research and technological approaches, ranging from nascent solutions to those that have received community adoption, whereas the opportunities take the form of open research questions that need to be explored.

The remainder of the paper is structured as follows: we start by outlining our motivating use case scenario (Section~\ref{sec:use-case}) and presenting the relevant background (Section~\ref{background}).
A conceptual framework is then proposed (Section~\ref{sec:conceptual-framework}), accompanied by an instantiation based on our use case scenario (Section~\ref{possible-instantiations}), after which we subsequently identify several challenges and opportunities (Section~\ref{challenges-and-opportunities}).  We then conclude by proposing a research roadmap for the governance of autonomous agents on the web (Section~\ref{conclusion}).

\section{Motivating Scenario}\label{sec:use-case}

\begin{figure}[t]
    \centering
    \includegraphics[width=0.6\linewidth]{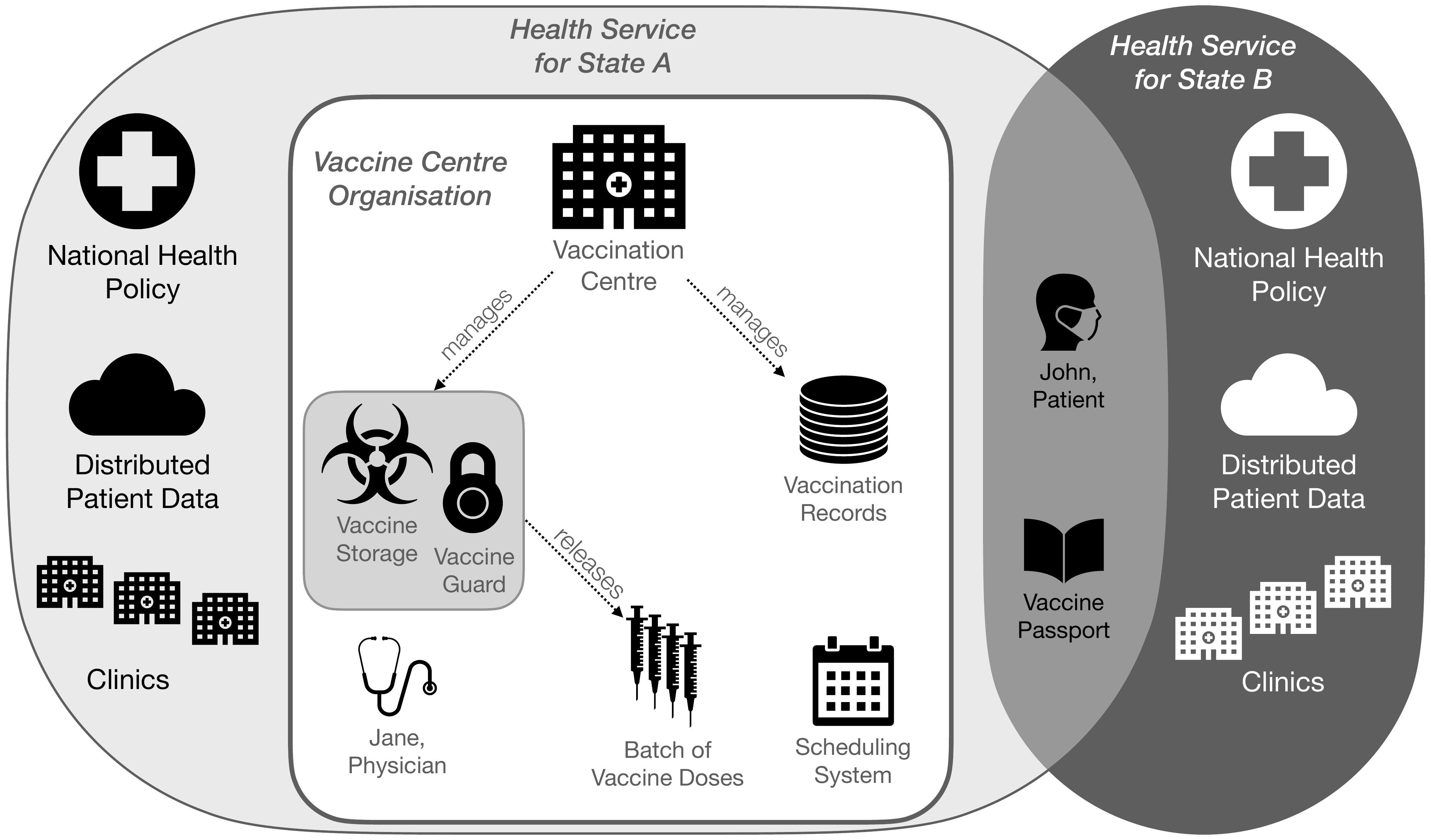}
    \caption{Organisations, agents, things, and services in the scenario.}
	\label{fig:use-case}
\end{figure}

The synergy between autonomous agents that leverage Web technologies, and the resources (i.e., things, services, information) that they can exploit
to achieve their goals can be illustrated through a motivating scenario that demonstrates the need for governance through the use of norms, policies, and preferences. 
Consider a scenario whereby there is a vaccination roll-out (for example, for the COVID-19 pandemic), where patients who request vaccinations may have differing personal circumstances. For example, John, the patient in Figure~\ref{fig:use-case}, may ask to be vaccinated early as he is the care giver for a vulnerable member of his family. As the demand for vaccines outstrips supply,  policies exist that determine vaccination eligibility. Furthermore, as vaccines are available from different manufacturers (e.g., AstraZeneca and Pfizer-BioNTech) and can be of different types (e.g., mRNA or inactivated vaccines), these vaccination policies may vary depending on the recipient's personal health record and/or their preferences, as well as vaccine availability.

Patients may be registered to different clinics or health centres that follow local or national policies or guidance on health care. In this case, John is registered at a clinic in his country (labelled State B in Figure \ref{fig:use-case}), but has a preference for vaccination near his current residential address in State A. Each country or state can be seen as having an \emph{organisation} of different health centres (clinics, hospitals, and vaccination centres), following their own national health policy that prescribe a specific specification/format for patient medical records, which may be held under disparate data models and access policies. Patient medical records are available (subject to appropriate authorisation) via web services using secure protocols across the web infrastructure \cite{TANWAR2020102407}, and are encoded using established medical ontologies and vocabularies to facilitate record exchange within and across different national health organisations.

Vaccination centres store batches of vaccines within one or more temperature-controlled vaccine storage systems, where each storage system is responsible for both inventory management and the dispensation of the different COVID-19 vaccine batches from a specialised cold store via a robotic arm. The release and retrieval of vaccine batches is guarded by policies that must be satisfied to ensure appropriate use by authorised personnel (i.e., the vaccine guard in Figure~\ref{fig:use-case}).
Once a batch of vaccines has been released, the vaccine doses should be used within a given time-frame to avoid spoilage and wastage, as they have a short shelf-life once thawed.
Furthermore, a scheduling system determines which patients can be vaccinated in a given time-slot, based on vaccination demand and patient requirement (determined by the current vaccination policy that may change frequently). 
This scheduling system should ensure that no vaccines are wasted, whilst ensuring that the policies determining which patients can receive which vaccines is adhered to.
Thus, the vaccination centre could be considered as an organisation that coordinates and exploits a variety of disparate information technology (IT) systems integrated through a Web infrastructure, including data management, scheduling, patient-facing services, and IoT-based physical assets such as the robot arm and the automated vaccine stores. Typically, however, the task of orchestrating and using these different systems requires costly and time-consuming human intervention.
Finally, once a vaccine has been administered, the patient's medical records should be updated, and the patient should be able to prove their vaccination status if required (e.g., using a vaccine passport \cite{9105054}).  The vaccine records should ideally be resilient to forgery whilst being privacy preserving and easy to administer \cite{9105054}; thus they may utilise a passport mechanism that itself exploits web-based resources such as \emph{verifiable credentials}\footnote{https://www.w3.org/TR/vc-data-model/}, decentralised data platforms~\cite{verborgh_iswc_2018}, blockchains~\cite{TANWAR2020102407}, etc.

This scenario raises challenges due to the decentralised and dynamic characteristics of the involved organisations, policies, services, and stakeholders. Patients can request vaccination based on their interpretation of eligibility, which should then be validated by the vaccination centre. The handling of requests may require the collection of patient data from multiple sources and the mapping to a shared data model. The vaccination eligibility policy can change frequently due to, for example, the emergence of a new \emph{variant of concern}, that may accelerate the need for vaccinating a specific population cohort or demographic. Changes to vaccination administration guidance may prioritise the use of certain types of vaccine over others for specific sub-groups (e.g., prioritising Pfizer-BioNTech over AstraZeneca, where possible, for certain patients based on medical risk assessments, or prohibiting certain vaccines for users where safety data is not available). Thus, the verification of vaccination eligibility for patients may rely on the aggregation of multiple policies, and on resolving inconsistencies between them.
A further challenge involves ensuring that the process for adhering to the national prioritisation criteria is fair and transparent.

Additional legal and ethical challenges arise when considering the complete socio-technical system, including electronic health record access \citep{10.1145/3365664} and supply chains \citep{5331908}.
Finally, vaccination scheduling needs to take into account patient availability (to avoid no-show cases and thus avoid vaccine wastage), as well as stock availability. Scheduling is therefore a collaborative process involving factors such as the vaccination centre capacity, vaccine availability, and patient availability. However, availability data may be distributed across multiple sources and, for privacy reasons, cannot be held centrally.

This scenario underlines the need for systematic and scalable approaches for the governance of the different IT systems and IoT-based physical assets, taking into account the need to operate under different governance institutions, as well as interact across organisational boundaries (e.g., between countries). Such interactions must comply with applicable norms and policies encountered at different stages of the vaccination roll-out. For example, the
European Commission recently proposed a Digital Green Certificate, recognised by all EU member states,  that facilitates the safe free movement of citizens within the EU during the COVID-19 pandemic.\footnote{\url{https://ec.europa.eu/info/live-work-travel-eu/coronavirus-response/safe-covid-19-vaccines-europeans/covid-19-digital-green-certificates_en}}

Given the intrinsic openness of the Web, coupled with the fact that autonomous agents can act on behalf of both patients and medical practitioners that need access to critical medical applications, the need for regulation, security, and privacy are of utmost importance. Additionally, there is a need to facilitate coordination between stakeholders and ensure that relevant regulatory requirements are adhered to throughout.

\section{Background}\label{background}

The vaccine administration scenario detailed in Section \ref{sec:use-case} relies on the availability of a uniform access layer that sits on top of several different systems (e.g., data management, services, and IoT platforms). The Web provides the necessary infrastructure to integrate and make accessible all of these systems, effectively becoming an application architecture for the proposed MAS~\cite{ciortea:emse-02070625}, on top of which autonomous agents may interact and cooperate to achieve common goals. In the following subsections, we present the relevant background in multiagent systems, the Semantic Web, and the WoT, followed by a discussion of the related work in norms, policies and preferences, with a focus on the governance of autonomous agents, both within and spanning those communities.

\subsection{Multiagent Systems}\label{sec:emas}

A multiagent system is composed of a (dynamic) set of \emph{agents} interacting inside a shared, possibly distributed, \emph{environment} which itself comprises a dynamic set of \emph{artefacts}. 
Agents are goal-oriented autonomous entities, encapsulating a logical thread of control, that pursue their tasks by communicating with other agents and by perceiving and acting upon artefacts within the environment. In essence, a MAS addresses the challenges of how agents may coordinate their efforts and cooperate in light of their autonomy \cite{weiss1999multiagent}.
Artefacts model any kind of (non-autonomous) resource or tool that agents can use and possibly share to achieve their goals. An agent perceives the observable state of an artefact, reacts to events related to state changes, and performs actions that correspond to operations provided by the artefact's interface. The coordinated and organised activities taking place in the system result from the concurrent and complex tasks handled by \emph{groups of agents} interacting with each other, or acting within an environment. Such activities may lead to recurrent patterns of cooperation captured by agent \emph{organisations}. Changes in the state of the environment may also lead agents to react and possibly affect the state of the organisation.

Research into multiagent systems has led to a number of concrete programming models.\footnote{Refer to the proceedings of the \href{http://emas.in.tu-clausthal.de/}{EMAS} or \href{http://paams.net/}{PAAMS} series for broad overviews.} These models\footnote{The models presented here reflect the relevant state of the art with respect to different MAS and are by no means exhaustive.}%
are concerned with agent-oriented programming \cite{Bordini-et-al07:book}, interaction and protocol languages \cite{AAMAS-BSPL-11}, environment infrastructures \cite{Weyns+07:environment}, and agent organisation model and management systems \cite{Ferber-et-al2004}. The results produced so far have clearly demonstrated the importance of these concepts and abstractions for the development of multiagent applications. Additionally, 
a variety of languages, tools, and platforms for agent-oriented programming (MAOP) have been developed and application success stories exists (e.g.,~\cite{DelaPrieta2019}). This type of research is often referred to under the umbrella of \emph{Engineering Multiagent Systems (EMAS)}. An overview and a comparative analysis of several prominent MAOPs can be found in \cite{kravari2015survey}.
One of the most prominent underlying architectures used by many agent-oriented programming systems is the \emph{Belief-Desire-Intention (BDI)} architecture, which models: knowledge (i.e., \emph{beliefs}) that the agent knows about, either through observation of the environment or interaction with other agents; goals (i.e., \emph{desires}) that the agent would like to bring about; and goals and plans of action (i.e., \emph{intentions}) that the agent is currently focused on.

From an agent development environment perspective, the 
Jade platform \cite{10.5555/1197665} provides a variety of behaviours (one-shot, cyclic, contract net) and is still available, although the last release dates back to 2017.  Although Jade does not directly provide support for BDI-based agents, they can be added through extensions such as Jadex \cite{10.1007/3-7643-7348-2_7}. Jack \cite{Jack-agentlanguage} is an example of a closed source BDI architecture, whereas the practical Agent Programming Language (2APL) is another open source language that retains BDI semantics \cite{DBLP:journals/aamas/Dastani08}. GOAL \cite{2a97d27ac7c4445aa6c877259e55d09d} offers a further BDI architecture which is actively maintained, whereas SPADE\footnote{\url{https://spade-mas.readthedocs.io/en/latest/index.html}} is a recently introduced Python-based BDI platform.
The JaCaMo MAOP framework,  based on the JaCaMo conceptual meta-model~\cite{boissier2020multi}, offers first-class abstractions to program the agents working environment and their organisation, in addition to offering the Jason interpreter for the BDI-based \emph{AgentSpeak} language \cite{Bordini-et-al07:book}.

Whilst MAOP is thriving within the academic community, industrial adoption of MAOP technologies is in its infancy, and standardisation efforts such as FIPA~\citep{FIPA-IP} (that superseded KQML) have received little attention in recent years~\cite{engineering-gsi-article-2019}.

\subsection{Agents and the Semantic Web}\label{sec:sw}

Attempts to tightly integrate autonomous agents and Web technologies date back to the vision of the \emph{Semantic Web} of the early 2000s. \citet{berners2001semantic} originally envisioned ``a web of data that can be processed directly and indirectly by machines'', in which intelligent agents act on behalf of humans, by searching for and understanding relevant information published on the web or acquired via services. Such information could potentially be made available by multiple sources, using alternative ontologies, often with different provenance. 
Autonomous agents rely on communication languages and protocols to exchange data and coordinate their behaviour and thus collaborate.
Early approaches based on speech acts \cite{austin1962how}, focused on message types or \emph{performatives} (e.g., \emph{request}, \emph{inform}, and \emph{promise}) based on a folk categorisation of the intended meaning of the communication. This evolved through the DARPA funded Knowledge Sharing Effort (KSE) resulting in a communication language, the \emph{Knowledge Query Manipulation Language (KQML)}, defining the mechanism by which agents communicated; and an ontology language, the \emph{Knowledge Interchange Format (KIF)}, describing the knowledge that the performative referred to \cite{10.1145/191246.191322}. Although agents could perform services on behalf of their peers, discovered through capability registries \cite{decker-sycara-williamson97}, service invocation occurred as a by-product of requesting information. This contrasts with the notion of \emph{web services} and \emph{things}, which use web-based communication protocols, whereby the invocation of services could be requested explicitly (in a similar manner to calling methods or functions within a programming language) by providing the relevant input parameters, as data or knowledge fragments. 

The prominent view from a Semantic Web perspective is that multiagent systems operate on the Web through the provision of services, using HTTP as the de facto standard transport protocol. Additionally, the Semantic Web community have developed standards, protocols, vocabularies, ontologies, and knowledge representation formalisms to facilitate the integration of machine-processible data from diverse sources at scale, using the existing web infrastructure. As such, the two communities diverged due to different priorities, though there is increasing recognition \cite{ciortea:emse-02070625} that the Web is a natural application architecture for MAS and can support different types of interactions between agents and resources.

From a knowledge representation perspective, standards such as RDFS~\cite{RDF_Schema_W3C:14} 
and OWL~\cite{OWL_Overview_W3C:12} 
facilitate the representation of complex knowledge about agents, services, things and their relationship in an explicit and processable way. An example is the Provenance ontology (PROV-O), a data model for workflows expressed using agents, their actions, and other assets.\footnote{https://www.w3.org/TR/prov-o/} Additionally, reasoning engines have been developed that are capable of reasoning over OWL ontologies, albeit often with some restrictions (cf.,  Pellet\footnote{https://www.w3.org/2001/sw/wiki/Pellet}, HermiT\footnote{http://www.hermit-reasoner.com/}, FACT++\footnote{http://owl.cs.manchester.ac.uk/tools/fact/}, Racer\footnote{http://www.ifis.uni-luebeck.de/~moeller/racer/}, and RDFox\footnote{https://www.oxfordsemantic.tech/}).
However, the use of ontologically grounded annotations for services within agent communication pre-dates the Semantic Web \citep{IC-col-1:6,DBLP:journals/tkde/FenselM01}, and in some cases the Web itself \citep{Icicis-93}. Semantic Web service research exploited both F-Logic \citep{10.1145/210332.210335} as used by WSMO \citep{DBLP:journals/ao/RomanKLBLSPFBF05}, and DAML-S \citep{damls-iswc02} (based on the DARPA Agent Markup Language) which evolved into OWL-S \citep{10.1007/978-3-540-30581-1_4}. Other approaches to support service utilisation were developed using OWL, e.g., the OWL ontology for protocols, OWL-P \citep{AOIS-06:OWL-P}, or using federated service discovery mechanisms such as the semantically annotated version of UDDI \citep{paolucci:02}. These frameworks and ontologies were key in facilitating the discovery and use of services by autonomous agents, and provided an alternative communication paradigm built on web-based infrastructure.
In addition, from the knowledge perspective, bespoke protocols were developed to support the decentralised management and exchange of knowledge and information amongst networks of agents or peers~\citep{DBLP:books/daglib/0014626}.

Other efforts include the provision of infrastructures for supporting the cleaning and validation of the data published on Linked Open Data Platforms; e.g., LOD Laundromat~\citep{10.1007/978-3-319-11964-9_14}\footnote{\url{https://github.com/LOD-Laundromat/LOD-Laundromat}} and OOPS~\citep{poveda2014oops}.\footnote{\url{http://oops.linkeddata.es}} Such techniques help detect errors in the data exchanged between agents and things. 
The SPARQL~\cite{SPARQL_W3C:13} query language facilitates federated querying over distributed data sources accessible via the web, whereas the Linked Data Platform~\cite{LDP_W3C:15} can be used to manipulate RDF data via  HTTP operations. Approaches have also been proposed to enrich SPARQL with qualitative and quantitative preferences~\cite{gue-pol-mci-iswc13,DBLP:conf/ekaw/Patel-Schneider18} to select query results that satisfy user-defined criteria.

In recent years the Semantic Web community has broadened its focus beyond knowledge representation, reasoning, and querying to include knowledge extraction, discovery, search, and retrieval. However, many of the proposed tools and techniques have yet to be used extensively within MAS or by the MAS community. A recent survey \cite{kirrane2021intelligent} identified several open research challenges and opportunities in relation to the suitability of existing proposals for autonomous agent use cases, the combination of symbolic and sub-symbolic AI techniques for enhancing agent learning, and the development of tools and techniques for validation and verification.

\subsection{Agents and the Web of Things}
\label{sec:wot}

The \emph{Web of Things (WoT)} \citep{W3C_WoT_Arch:20} refers to the Internet of Things (IoT) with an application of Web standards and technologies for improving interoperability of IoT devices and infrastructure. 
Things are resources that can be acted upon or queried via APIs (e.g., WoT scripting API~\citep{wot-scripting-api:20}); \emph{autonomous goal-driven agents}\footnote{Here, we refer to agents in the sense of multiagent systems as discussed in Section~\ref{sec:emas}.} thus can make use of a WoT environment via WoT technologies and become part of the WoT ecosystem.
Indeed, bringing agents to the Web requires more than simply exploiting Web protocols (such as HTTP \citep{HTTP_RFC:14}) and data formats (e.g., XML~\citep{XML_W3C:08}, RDF~\citep{RDF_Concepts_W3C:14}). The communication infrastructure used by agents should comply with an architectural style based on well-defined principles, such as \emph{Representational State Transfer (REST)}~\citep{Filding_REST:00} as instantiated in the Architecture of the World Wide Web~\citep{WebArch_W3C:04}.\footnote{Note that the specification of the Web architecture defines the concept of \emph{Web agents} as ``a person or a piece of software acting on the information space on behalf of a person, entity, or process''.} 
Furthermore, for things to be used without human intervention, they must be formally described. To this end, the W3C published the \emph{Thing Description}~\citep{TD_W3C:20} standard, which specifies how a JSON-LD representation of thing affordances (i.e., properties or actions) via Web APIs can be provided. 
In addition, the \emph{WoT Discovery}~\citep{WoT_Discovery_W3C:20} standard provides a mechanism for the automatic discovery of thing descriptions (thus obviating the need to hard-code the location of such descriptions beforehand). 
These standards support improved heterogeneity by decoupling agents from thing implementation details.

The WoT activity highlights the importance of metadata with clear semantics, and made their standards, especially thing descriptions, compatible with RDF and Semantic Web technologies. In fact, even before a standardisation effort for the WoT started, multiple initiatives suggested the use of the Semantic Web to improve IoT systems~\cite{DBLP:journals/iot/RhayemMG20}. More precisely, in REST style hypermedia systems such as the WoT, things and agents are resources that interact by producing and consuming hypermedia about their state and the artefacts surrounding them \cite{ciortea:hal-02196903}. All resources are identified through IRIs\footnote{Internationalised Resource Identifiers~\cite{IRI_RFC:05}
} to support global referencing, irrespective of contextual information. Therefore, resources can be represented through semantic descriptions that are expressed in a uniform data exchange format such as RDF %
using terms from some standardised and interlinked vocabulary expressed in OWL~\citep{OWL_Overview_W3C:12}. This standardised knowledge model hides the specifics of the implementation and facilitates interconnected resources that can be queried by exposing SPARQL endpoints. Of particular interest to WoT environments are the vocabularies that describe sensors and actuators (SOSA/SSN~\cite{SSN_W3C:17}), provenance (PROV~\cite{PROV_O_W3C:13}), and temporal entities (OWL-Time~\cite{OWL_Time_W3C:17}).

The WoT provides a natural substrate for multiagent systems based on the vision that systems of interconnected things should be open and easily reconfigurable, and therefore such systems should comprise autonomous and collaborative components. This notion was supported by
\citet{DBLP:conf/icdcs/SinghC17} who argue that IoT systems need the kind of decentralised intelligence that MAS provides. Likewise, \citet{ciortea:emse-02070625} recommend integrating the Web and MAS to leverage the proven benefits of hypermedia systems for MAS. Importantly, these papers emphasise governance as a major challenge.

The technologies that emerge from the WoT community are often industry-oriented and paralleled by standardisation efforts.
A recent example is the abstract \emph{WoT architecture} design document~\cite{W3C_WoT_Arch:20},
supported by the \emph{Thing Description}~\cite{TD_W3C:20} and the \emph{WoT Scripting API}~\cite{wot-scripting-api:20} specifications, for which a reference implementation is provided.\footnote{\url{https://github.com/eclipse/thingweb.node-wot}}
Although these technologies are more mature than MAOP technologies from an engineering perspective, and have a clear path to industry adoption, they lack the rich abstractions related to agents and autonomy that MAOP technologies provide.
For example, the notion of a \emph{servient}, as introduced in the WoT architecture design document can be considered an evolutionary step from a stricter server-client separation; a notion that is considered simplistic within the MAS community.
Recent approaches have sought to form a bridge between the MAOP and WoT technology ecosystems~\cite{10.1007/978-3-319-91899-0_8,DBLP:conf/emas/CiorteaBR18}; however, this line of research is young and the corresponding technologies are nascent.

\subsection{Norms, Policies and Preferences}

Norms, policies, and preferences can help govern autonomous agent behaviour. The term \emph{norm} has several meanings in natural language and is used widely in economics and social science. In MAS, the term ``norm'' typically expresses a deontic concept (e.g., a prohibition, permission, obligation, or dispensation). A coherent set of norms, i.e., created and evaluated as a unit, is referred to as an institution~\cite{noriega-thesis}. 
The same understanding of norms is found in the Semantic Web literature, where there is also a body of work focusing on policy specification and enforcement.
Here, \emph{policy} is an overarching term used to refer to a variety of system constraints,  whereas the term preferences is primarily used in connection with privacy and personal data protection.

The study of norms is a long-running and active line of research within the MAS community, as evidenced by numerous Dagstuhl seminars~\citep{dastani2018normative,andrighetto2012normative}, and a handbook on the topic~\citep{1258215}.
Normative MAS~\citep{boella2006introduction} are realised and characterised in multiple ways, including those based on: (1) the agents reasoning capabilities; (2) whether norms are implicit or explicit; and (3) whether or not the architecture includes monitoring and enforcement mechanisms.

Agent capabilities vis-à-vis norms typically fall into three categories:
\begin{inparaenum}[(i)]
\item \emph{norm unaware}, whereby agents may be regimented by external agencies to enforce norm compliance~\cite{DBLP:journals/eaai/ArcosENRS05};
\item \emph{norm-aware}, where agents may choose whether or not to comply with norms, depending on the alignment of their goals with those norms, the penalties for non-compliance, and the likelihood of enforcement~\cite{DBLP:journals/taas/ShamsVOP20}; and
\item \emph{value aware}, whereby agents, in addition to being norm-aware, are able to participate in norm creation and norm revision, by reasoning about the values supported (or not) by particular norms~\cite{DBLP:conf/ijcai/CranefieldWDD17}.
\end{inparaenum}
Thus, compliance in normative systems depends on how individual agents reason and adapt to norms at both design and run time~\cite{Riemsdijk-et-al2013,JAIR-19:goals+commitments,DBLP:conf/dalt/LeePLDA14}.
Implicit norms that reside within the agents themselves are expressed through agent behaviour, but are not otherwise externally discernible, whereas explicit or referenceable norms may have an abstract representation involving variables and a grounded (detached) representation in an entity such as a contract~\citep{10.1145/2542182.2542203},  institution~\citep{noriega-thesis,10.1007/3-540-45448-9_26,DBLP:journals/ai/dInvernoLNRS12}, or organisation~\citep{AAAI-VO-06,boissier2020multi}.
Agents that are norm or value aware should be able to:
\begin{inparaenum}[(i)]
\item recognise norms; 
\item decide whether they want to follow them; and
\item adapt their behaviour according to the norms, if they decide to do so.
\end{inparaenum}
Such agents may additionally be able to engage in norm revision processes.
Norms, and more broadly \emph{conventions} or \emph{social norms}~\cite{lewis:69a}, are established in an agent society in one of two ways, namely top-down and bottom-up~\cite{wooldridge2001:intro,normEmergence}.

In \emph{top-down} systems, norms are identified as part of the MAS design process and are either: hard-coded into the agents' behaviour (implicit representation), eschewing any form of normative reasoning and narrowing the scope for behavioural adaptation; or are prescriptive and explicitly represented, and thus external to the agents, typically represented in the form of abstract regulations (for example, ungrounded terms over variables) that, as a result of agent actions, become detached (for example, grounded terms over literals). The n-BDI variant \cite{criado2010towards} is a BDI-based agent architecture that allows for the internalisation of norms where the design suggests an agent-internal process that synthesise norm-style rules based on observed behaviour, whereas N-Jason~\cite{DBLP:conf/dalt/LeePLDA14} agents perceive institutional facts, which they internalise as beliefs and hence incorporate in their reasoning. Norms designed offline, however thoughtfully crafted for the long-term, are at risk of losing relevance in open, always-on, environments such as the Web, because it is not possible to anticipate all eventualities at design time. Furthermore, drift in the agent demographic or in systems goals, are likely to make norm revision essential over any sufficiently long system lifetime. With explicit norms, any norm change will affect the entire population. Such changes can be effected through a human-in-the-loop approach, where human designers revise the norms and then switch the system over at some suitable point;  such as through a shutdown/reboot sequence, or the use of norm-aware planning~\cite{DBLP:journals/taas/ShamsVOP20}.  In the latter case, an agent must manage a plan sequence that although initially compliant, may cease to be part of the way through the plan due to the change in norms.  Such an agent must also be able to check that its learned way of achieving a goal is compliant with the new norms, perhaps by means of some oracle~\cite{padget-et-al:2018}, or by being able to acquire a fresh plan that is compliant.

In \emph{bottom-up} systems, an individual agent decides whether or not to adopt a norm: with implicit norms, it may seek advice from others or apply indirect reinforcement learning over its observations, as a basis for prediction, possibly in combination with a \emph{strategy update function}~\cite{wooldridge2001:intro}. In such systems, norms are deemed to have emerged once they have been adopted by a sufficiently large fraction of the population; this is  typically 90\% in most of the literature, and 100\% in some cases (which is hard to achieve), or assumes a simple majority, which can risk oscillatory outcomes. However, convergence (this term appears to be used interchangeably with emergence in the literature~\cite{morris-martin-et-al:2021}) is a function of the \emph{capabilities} of the agents. Emergence with explicit norms depends on agent reasoning capabilities. An agent might inform the regulator that it wants to take a particular action in a particular state (without sanction)~-- the agent knows what it wants but not how to get it---as a request to change the norms without having to reason about norm representation. A more difficult approach is that an agent might propose a new (abstract) norm~-- the agent knows how to define a new norm to get what it wants~\cite{haynes2017engineering,morris-martin-et-al:2021}. As above, changes have to be actioned, which could be as outlined previously, although pluralist approaches are possible, as put forward by \citet{ostrom:1990}, or by using one of the many voting mechanisms. The challenge for an agent then becomes how to decide which way to vote, which depends on their reasoning capabilities: are they able to evaluate the consequences of the norm change; and are they selfish (i.e., vote ``yes'' if the change is individually beneficial, e.g., increases their utility) or altruistic? (i.e., vote ``yes'' if the change is collectively beneficial). More sophisticated still would be the use of argumentation to determine if the revision is consistent with the population's values \cite{DBLP:journals/taas/ShamsVOP20,10.1007/978-3-319-08615-6_17}.

In the early days, Semantic Web researchers proposed general policy languages, such as KAoS~\citep{bradshaw1997software}, Rei~\citep{Kagal2003} and Protune~\citep{Bonatti2007}, which cater for a variety of different constraints (access control, privacy preferences, regulatory requirements, etc.). A prominent early attempt to provide a semantic model of polices as soft constraints for agents was OWL Polar~\cite{SENSOY2012148}, an OWL DL explicit policy representation language. OWL Polar aims to fulfil the essential requirements of policy representation, reasoning, and analysis, where policies are system-level principles of ideal activity that are binding upon the components of that system, and thus are used to regulate the behaviour of agents~\cite{SENSOY2012148}.
Over the years the Semantic Web community have also proposed policy languages that are tailored to better cater for access control, privacy preferences, licensing, and regulatory governance requirements,
including detailed surveys, for example, of the various policy languages, and the different access control enforcement strategies for RDF \cite{kirrane2017access}.
From a privacy perspective, the Platform for Privacy Preferences Project (P3P)~\cite{world2002platform} specification, deemed obsolete in 2018, aimed to allow websites to express their privacy preferences in a machine readable format that could be interpreted by agents that could automate decision making on behalf of humans. The P3P initiative, despite having failed, inspired subsequent work on representing and reasoning over privacy preferences, such as
using OWL~\cite{garcia2008web}, catering to more expressive
privacy preferences~\cite{kolari2005enhancing}, and representing consent for personal data processing~\cite{bonatti2020machine}.

Many existing proposals rely on WebID~\cite{W3C_WebID:14}, a community-driven specification that offers an identification mechanism making use of Semantic Web technologies to provide password-less authentication. An extension of WebID (specifically WebID-OIDC that relies on OpenID Connect\footnote{\url{https://openid.net/connect/}}) is used in the \emph{Solid} project. Solid\footnote{\url{https://solidproject.org/}} is an ongoing initiative, lead by Tim Berners-Lee, aimed at deploying a distributed Linked Data infrastructure for governing one's personal data, which is built on top of Linked Data Platforms. 
Additionally, there has been work on \emph{usage} control in the form of licensing~\citep{cabrio_these_2014,villata2012licenses,guido_heuristics_2013,governatori2013one,governatori2014live}, and more recently, policy languages have been used as a means to represent regulatory constraints~\citep{palmirani2011legalruleml,de2019odrl}. The Open Digital Rights Language~\cite{ODRL_W3C:18,fornara2019}, although primarily designed for licensing, has been extended to cater for: access policies~\cite{steyskal2014}; requests, data offers and agreements \cite{steyskal2015if}; and regulatory policies~\cite{de2019odrl}. Usage control, however, often proves challenging for organisations and users, and any constraints imposed on the use of data need to ensure that policies are applied consistently across organisations and that there are robust propagation mechanisms preventing policies from becoming invalid~\citep{10.1145/2815833.2815839,7580764}.
The notion of FAIR ICT Agents \cite{kirraneintelligent} is based on FAIR (Findable,
Accessible, Interoperable and Reusable) principles \cite{wilkinson2016fair}, where ICT denotes \emph{interactive intelligent agents that are constrained via goals, preferences, norms and usage restrictions}.
Thus far, the WoT standards offer only limited support for norms, policies and preferences, which are currently described in  guidelines targeted at human developers rather than as declarative, machine-readable statements usable by agents~\citep{W3C_WoT_Security:19}.

Although research on norm-aware agents has made reasonable progress to date, much remains to be done to elevate human oversight to align with the three categories \cite{aihleg}: \emph{human-in-the-loop}, where there may be human intervention in each decision cycle; \emph{human-on-the-loop}, where there is human intervention in the design cycle and operation monitoring; and \emph{human-in-command}, where there is human oversight of the overall system, including the means to decide when and how to engage the AI system. The motivated scenario presented herein draws on human-on-the-loop and human-in-command, and indeed it is these levels of abstraction that inspire the governance framework introduced in Section~\ref{sec:conceptual-framework}, since those are the system characteristics we aim to facilitate.

\section{Conceptual Framework}
\label{sec:conceptual-framework}
\begin{figure}[b]
    \centering
    \includegraphics[width=0.45\linewidth]{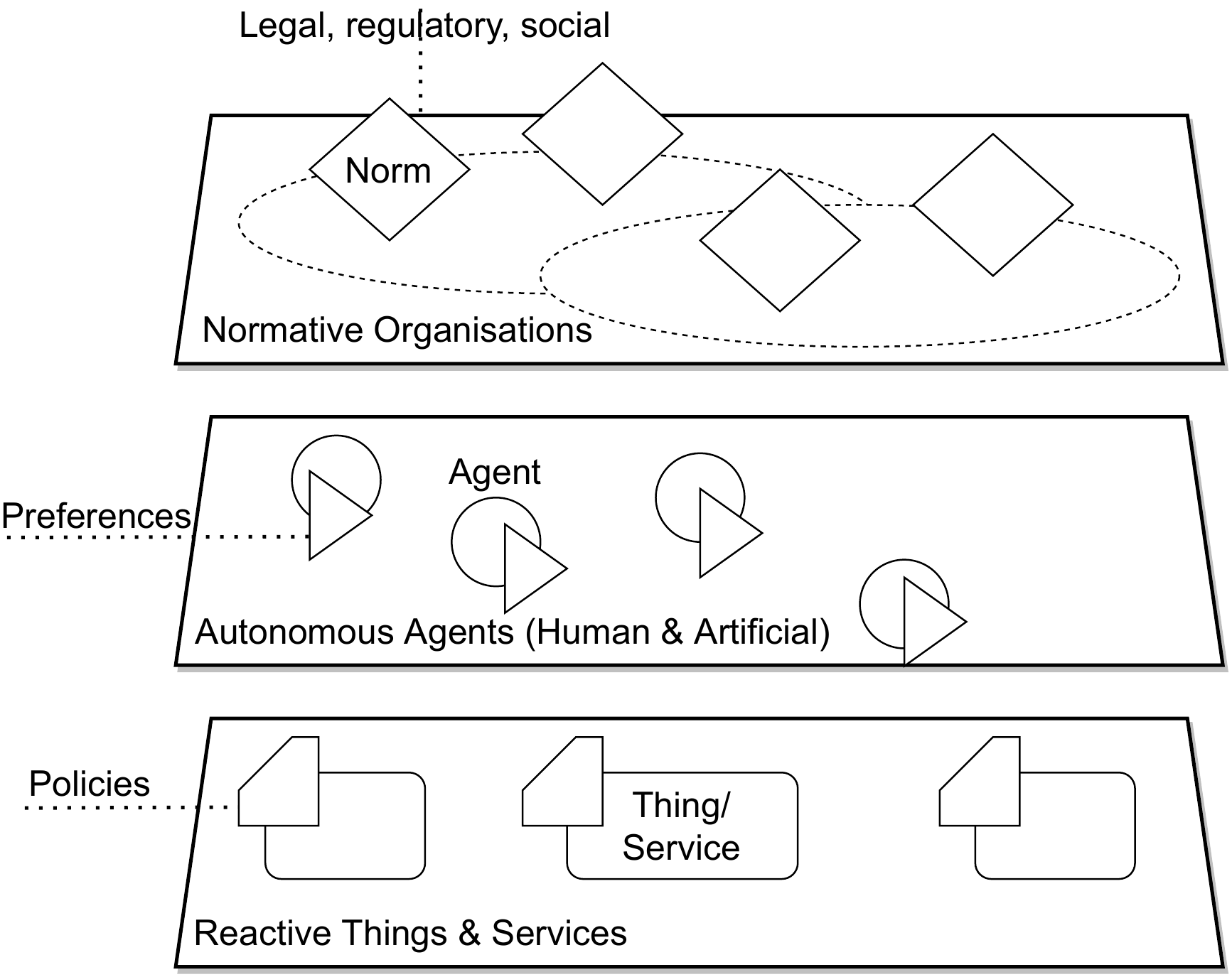}
    \caption{Conceptual Framework for Governing Agents on the Web structured along three layers. Interactions (not represented here in this Figure) take place within each of the layer and between the layers.}
	\label{fig:framework} 
\end{figure}

The overarching goal of this section is to identify governance entities, their relations and their purpose, with no aim to be prescriptive in their instantiation. In doing so, we propose a blueprint for the governance of socio-technical systems that can be instantiated in a variety of ways, using a variety of concrete software components. Thus, this section aims to provide guidance for developers on how different parts of an agent governance system fit together and the functions that they contribute. Our objective here is to enable a range of solutions, fit for different purposes, realisable through available (rather than prescribed) software, but still coherent through the framework set out in the three layers shown in Figure~\ref{fig:framework}.

\subsection{Global View}
\label{sec:conceptual-framework:global}

In order to provide something actionable for designers and implementors, we ground our framework for the governance of autonomous agents on the Web through three layers that structure the various entities and abstractions needed for the development of socio-technical systems on the Web.  Each layer is assigned concepts that are necessary for governance: norms, policies, and preferences (as illustrated in Figure~\ref{fig:framework}). 
The way in which these different parts are realised, and how they interact is dependent on various design decisions.
In setting out this framework, we draw on and organise existing work on norms, policies, and preferences (as described in Section~\ref{background}) to cater for abstract requirements for the governance of socio-technical systems. This gives rise to the following three layers:

\noindent \textbf{Reactive Things \& Services Layer.} This layer comprises non-autonomous entities in the environment.
 As seen in Section~\ref{background}, such entities are key notions of the WoT architecture~\cite{WebArch_W3C:04} for which first-class abstractions are proposed for specifying and discovering them and other entities within the MAOP approach (e.g., artefacts in the JaCaMo meta-model~\cite{boissier2020multi}). Such entities are perceived and acted upon by agents.
 We propose the use of \emph{policies} for dealing with the governance of such non autonomous entities, following the same approach adopted by Semantic Web community. These policies state who can access them, and constraints on their usage (if any). 
Semantic Web technologies such as OWL-POLAR~\cite{SENSOY2012148} can provide a means to implement, manage, and enforce policies that constrain access to things and services, and the affordances they provide.
\noindent \textbf{Autonomous Agents (Human \& Artificial) Layer.} The agents layer is composed of entities that autonomously perceive and act upon their environment (i.e., things and services) and interact with the other entities.
Agents are the main abstractions for specifying and managing autonomous behaviours. In contrast to the conventional model of programs on the Web as \emph{servers} or \emph{clients}, the WoT architecture introduces a \emph{servient} that can both pro-actively access other things and services and reactively respond to requests from other things and services. In addition, servients can host one or several things. Whilst the Web architecture does not provide first-class abstractions for autonomy, it is possible to
distinguish between ``agentified'' things that exhibit pro-active behaviours and reactive things by introducing custom properties into the W3C WoT \emph{Thing Description}~\cite{10.5555/3463952.3464231}.
Agents have \emph{preferences} that inform and constrain their actions with respect to things, web services, and other agents.
Preferences control the local reasoning and decision-making undertaken by the agents, and can thus support governance.
In traditional deliberation architectures for autonomous agents, preferences are specified (or emerge) as part of the often complex reasoning cycles. Hence, the management of these preferences given the presence of norms and policies can be challenging. Semantic Web approaches that consider preferences (e.g., SPARQL with preferences~\cite{10.1145/2851613.2851690}) can enable declarative preference management, especially when an agent's preferences are to be considered.%
\noindent \textbf{Normative Organisations Layer.}  
In MAOP, organisations are first-class abstractions~\cite{boissier2020multi} that group agents and their governance (i.e., norms). Although the WoT architecture does not provide such abstractions, its security and privacy guidelines reflect similar notions to organisational norms. 
Whilst the previous two layers (discussed above) included governance concepts dedicated to the local governance of each entity (e.g., policies for thing, preferences for agent), this layer addresses the governance of autonomous entities \emph{participating in the system}.
This layer manages abstractions for the logical grouping of agents with a particular purpose, and the provision of legal, regulatory, and social norms that may possibly span multiple organisations. However, organisations are entirely virtual and passive (i.e., shaped by their members), thus it is up to these member agents to stipulate, comply with (or violate), enforce, and evolve organisational norms.
Semantic Web technologies such at ODRL~\cite{fornara2019} allow for the formalisation of norms for specific domains and purposes; hence, they can be integrated seamlessly with the more abstract MAOP abstractions for organisations and norms that are agnostic to these details.

From a MAS perspective, this framework is coherent with the JaCaMo meta-model~\cite{boissier2020multi}; from a WoT perspective, it is coherent with the WoT architecture~\cite{WebArch_W3C:04}; and furthermore, it is coherent with the Semantic Web perspective, although with enhancements with respect to policies, preferences, and norms.
It is worth noting that our conceptual framework provides \emph{software engineering abstractions}. Analogously to the Web architecture, we do not recommend a one-to-one mapping of software abstractions to physical entities (devices). Considering Web architecture standards, the WoT Scripting API supports, for example, the instantiation of multiple things as part of one \emph{servient}, which may represent a single physical machine.

\subsection{The Layers}
In this section, we return to each of the layers introduced in Section \ref{sec:conceptual-framework:global} and detail their composition and their governance. 

\subsubsection{Reactive Things \& Services}

As defined in Section \ref{sec:wot}, things are physical objects that are endowed with network capacities that allow one to make use of their functions in a digital environment. For example, they can be sensors that provide measurement data through the Internet, or actuators that can be triggered from a Web API. Therefore, in the WoT context, things resemble and are sometimes assimilated within web services. These web services are normally purely digital entities that simply exchange data via their input parameters and output results.
In the WoT architecture, things may be autonomous, whereas, in contrast, our conceptual framework distinguishes between autonomous agents (which may be things in the WoT architecture) and \emph{reactive} things.

When it comes to things and services, policies serve many purposes. Access control policies ensure that only authorised agents use specific things and services. Here, there is a need to provision both authentication and authorisation mechanisms, and policies may help resist security threats. Additionally, policies may govern the use of data that is produced by things and services; e.g., to ensure personal data protection or intellectual property rights. 

From a policy governance perspective, it is useful to distinguish between \emph{enforcement} and \emph{compliance} of the autonomous agents acting on these things and services given their respective policies. Enforcement means that any violations are prevented, whereas compliance means there is a need for retrospective conformance checking. 

\subsubsection{Autonomous Agents (Human \& Artificial)} 
In contrast to (reactive) things and services,
agents are entities that pursue their own goals autonomously. They determine
the necessary actions that should be executed on the things and services situated in the environment.
In the MAS literature, several agent architectures that are based on the different properties exhibited by the agents have been proposed~\cite{Wooldridge-Jennings95:KER}. They range from purely reactive (i.e., those that respond to stimuli without complex symbolic reasoning to reason about future actions) to deliberative ones (those that maintain a symbolic world model for reasoning about plans and decision making)~\cite{Calegari-et-al21:jaamas}. A notable example of deliberative architectures is the BDI architecture~\cite{Rao-Georgeff95:icmas}, where agents are programmed using their mental attitudes such as beliefs, desires, wishes, etc.~\cite{Bordini-et-al07:book} and that is one of the mainstream architectures for cognitive agents in MAS.\footnote{A number of different MAOP frameworks that adopt a BDI architecture were discussed previously in Section~\ref{sec:emas}.} This contrasts with reactive architectures (such as the subsumption architecture~\cite{BROOKS1991139,DBLP:journals/connection/TheodorouWB17}) typically used by robotic systems, whereby \emph{behaviours} define the actions a robot should perform as a consequence of some stimuli (e.g., from sensor data or direct communication). Many hybrid agent architectures~\cite{Wooldridge-Jennings95:KER,DBLP:conf/ijcai/BrysonS01} combine elements of both reactive and deliberative ones, where prominence is often given to the reactive aspect over the deliberative aspect (such as obstacle avoidance versus goal deliberation).
Our conceptual model focuses on governance and is agnostic with respect to any particular architecture, and thus cater for the heterogeneity of agents.

In addition to taking decisions on their own, agents may also coordinate with humans or with other agents to adjust and align their goals with the other agents' goals and identify joint goals, and as such, they may communicate with other agents or human users by exchanging messages. We address the various means of interaction among agents in Section~\ref{sec:interaction} below.

Each agent maintains a representation of its internal state that is built from the agent's internal reasoning, from its perceptions of the environment, i.e., the observable state of the things and services deployed in the system, and from its interactions with other agents.
Acting on behalf of human users (e.g., assistant agents) or abstract entities (e.g., service agents), agents manage preferences that guide their decision process. It is important to differentiate between agents developed by the application designer and those that enter the system at run-time. 
This differentiation emphasises the level of control the application designer has over the agent with respect to its internal state. It also justifies the proposition of two levels of governance within our conceptual model:
preferences for local and individual control; and organisations for global and collective control.
In our conceptual framework, preferences cover many dimensions, ranging from privacy preferences to moral values or ethical principles. 
Additionally, there can be either agreement or conflict between preferences and access control as defined in the previous layer, due to the fact that an agent may need to verify someone's identity, and based on this determine what information to disclose. Being part of a MAS, the reasoning and decision mechanisms of the agents are enriched with mechanisms to reason over several factors including: norms; regulatory requirements coming from the organisation in which the agent participates; and over policies or access control rules attached to the resources, things, and services with which the agent interacts.

\subsubsection{Normative Organisations}

Organisations act as coordination mechanisms by which agents work together to achieve their joint goals. The design of agents within an organisation focuses mainly on the agents' capabilities and constraints, as well as on organisational concepts such as \emph{roles} (or functions, or positions), \emph{groups} (or communities), \emph{tasks} (or activities) and \emph{interaction protocols} (or dialogue structures); therefore on what relates the structure of an organisation to the externally observable behaviour of its agents~\cite{Ferber-et-al2004}.
Organisations usually have a structure defined by: (i) groups, whereby agents are classed together and possibly organised hierarchically; and (ii) roles, whereby agents assume various duties. 
For example, agents can belong to multiple organisations, be part of various groups, assume different roles (possibly at the same time), and join or leave organisations at will. Organisations can be formed at design time or emerge due to interactions between agents at run-time. 

The dynamics of the organisational structure, for example an agent changing its role or joining a group, is governed by rules that are formalised as the norms of the organisation. Norms define what communication is possible, allowed, or forbidden between agents. An organisation is a means to regulate agent behaviour, and such organisations may be governed by norms, including laws and regulations adopted from the social setting or jurisdiction and those legislated within the organisation. The organisation structure and its normative part are described in such a way that agents can autonomously take part in the organisation and regulate themselves automatically with the aim of achieving their (individual or collective) goals. However, a formal, explicit encoding of norms is necessary to facilitate automated compliance and conformance checking.

\subsection{Interactions Within and Among the Layers}\label{sec:interaction}

Within a MAS that is fully aligned with our layered conceptual framework, several interactions may take place within each of the layers and across them (i.e., both inter and intra-layer interactions). 
Similarly to the Web architecture, the conceptual framework is protocol-agnostic. Some protocols may be chosen based on the underlying things and services to be used; whereas other protocols would be custom to the desired agent-agent interactions; and some of these latter protocols may be designed whereas others evolve. We identify the following types of interactions in the conceptual framework:

\noindent \textbf{Agent-to-agent interactions.}
    Agents can interact with other agents \emph{directly}, by exchanging messages or acting upon each other, or \emph{indirectly}, by observing each other's actions on the reactive things and services of their environment.
    Because agents are autonomous, the requests that one agent sends to another are handled at the discretion of the receiving agent. In comparison to the interaction with Web services, interaction with an agent may imply a higher likelihood that the response deviates in a complex and nuanced manner from the requested resource.\\
\noindent \textbf{Agent-to-thing/Thing-to-agent interactions.}
    Agents \emph{proactively} interact with things and services by \emph{acting} upon them, \emph{accessing} their properties, and by listening to (perceiving) events that things and services emit. \\
\noindent \textbf{Thing-to-thing interactions.}
    While things and services are purely \emph{reactive}, they may interact with other things or services as part of a reaction chain. In this context, existing standards that are part of the Web architecture can be applied for basic communication, but more expressive approaches may be required to manage norms, policies, and preferences, for example when a thing communicates on behalf of an agent across organisational boundaries. \\
\noindent \textbf{Agent-to-organisation/Organisation-to-agent interactions.}
    An agent's preference depends on the norms of the organisations that the agent is a part of. However, because the agent is an autonomous entity, it may choose to not adopt an organisational norm. At the same time, the agent may attempt to change an organisation's norm, for example by proposing a norm update that then requires approval by a majority of the organisation's agents. \\
\noindent \textbf{Thing-to-organisation/Organisation-to-thing interactions.}
    In contrast to agents, things and services cannot directly affect organisations. Things and services can be implemented to dynamically adopt policies that reflect organisational norms, and the state of a thing or service can be considered by an organisation, but in both directions, the organisation is the \emph{leading system}.\\
\noindent \textbf{Organisation-to-organisation interactions.}
    Several organisations may have (unidirectional or bidirectional) dependencies. For example, in a hierarchy of organisations, the norms of lower ranking 
    organisations may depend on norms that are specified on a higher level in the hierarchy; still, a higher ranking organisation may have some norms that depend on the norms of multiple lower ranking organisations (consider dependencies between a federated state and its federal entities).

In the MAS community, interaction protocols are typically designed from a global perspective and aim to facilitate interaction and coordination between agents. A protocol specifies the permitted enactments; i.e., the possible sequences of message exchanges. Proposals for languages for interaction protocols include process algebra~\cite{Ferrando+19:enactability}, Petri Nets~\cite{Reisig1985:book}, and information protocols \cite{AAMAS-BSPL-11}. Petri nets may then be mapped to models that are more accessible to human users, such as Business Process Model and Notation (BPMN) diagrams. In practice, protocol design and protocol discovery can go hand-in-hand: in particular, Petri Net-based protocols (\emph{processes}) can be \emph{mined} from IT system event logs~\cite{van2012process}, which, for example, can be used for organisational compliance checking~\cite{CARON20131357}. Recently, \emph{agent system mining} has been proposed as a novel process mining variant that focuses on the agents that participate in one or several (organisational) processes, i.e., on the micro-level instead of on the macro-level process view that an organisation imposes~\cite{tour2021agent}.
In the service-oriented community, the notion of a \emph{choreography} is similar to an interaction protocol~\cite{Barker-et-al2009:choreography}, in that a choreography describes interactions between services from a global perspective.

Human-Agent interactions are typically modelled as conversations between the different agents, i.e., \emph{dialogues}~\cite{walton1995commitment}. A dialogue has a normative aspect: it is regulated by norms, and can establish new norms. In normative systems, dialogue protocols are specific notations for norms that specify the violation contexts. Utterances in a dialogue can be seen as moves in the underlying protocol that create obligations and permissions for the participating agents. Of particular interest are persuasive dialogues, where an agent can \emph{convince}, \emph{suggest}, or \emph{command}. Agents can use persuasive dialogues to convince other agents to add new beliefs; to enter into some form of negotiation; or, in the case of the command, a new violation rule is introduced thus creating a new obligation~\cite{Boella2004e}.

Besides the protocols themselves and their logical organisation in process choreography, the Agent Communication Language (ACL) and the agreed vocabularies are crucial when it comes to the interaction and co-ordination of agents in a MAS (Section \ref{sec:sw}).
The Web architecture specification lists \emph{properties}, \emph{actions}, and \emph{events} as the central abstraction of its interaction model. In our conceptual framework, organisations, agents, things and services may expose properties and generate events, but only agents may execute actions.
From a governance perspective, there is a need for policy, preferences, and norm-aware interaction protocols. For example, agents may need to authenticate themselves to other agents as well as to things and services, whereas collaborating agents may need to engage in preference elicitation and negotiation, and norms may need to be communicated and possibly agreed upon by agents that form part of an organisational structure.

\section{Use Case Revisited}
\label{possible-instantiations}

In this section, we demonstrate how a \emph{normative} MAS that leverages web services, things, and Semantic Web technologies could be used to realise our motivating scenario (Section \ref{sec:use-case}). We show how several example situations can be modelled using the proposed conceptual framework (Section~\ref{sec:conceptual-framework}) and highlight technologies that could be used to instantiate our governance framework.

\begin{figure}[t]
    \centering
    \includegraphics[width=0.5\linewidth]{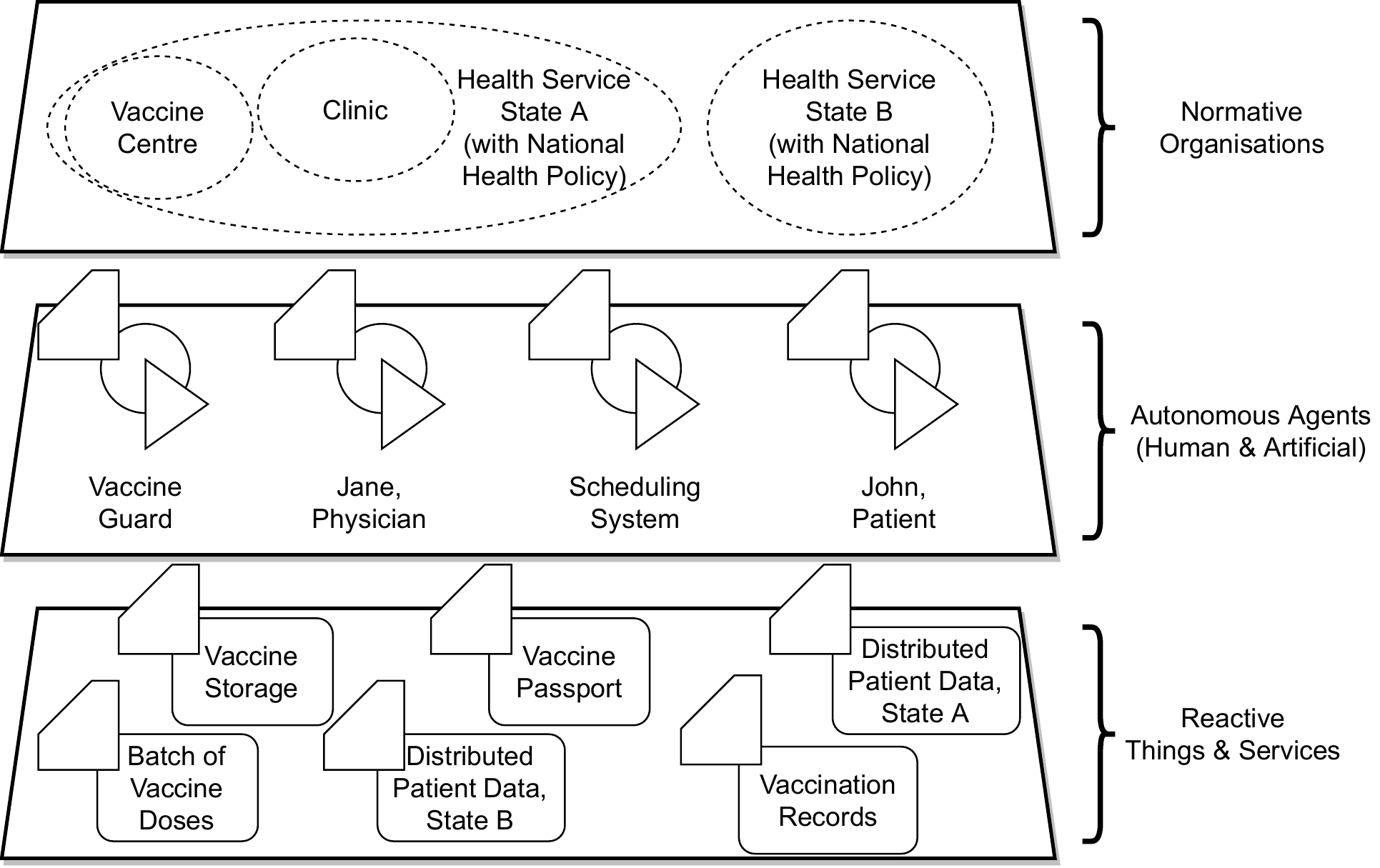}
    \caption{Normative organisations, autonomous agents, and reactive things and services in the scenario.
    }
	\label{fig:use-case-model}
\end{figure}

\subsection{The Global Setting}

Agents encapsulate knowledge, goals, and preferences corresponding to the autonomous entities involved in the vaccination process. The resulting conceptual model is illustrated in Figure~\ref{fig:use-case-model}. 

An assistant agent is in charge of managing personal data on behalf of a patient (e.g., the patient John). 
A physician agent is in charge of managing administrative tasks to act on behalf of the physician (e.g., Jane). 
Other types of agents access the things and services (i.e., a \emph{vaccine guard} agent controls access to the freezer), and to manage the vaccination process by collecting patients' data and checking their eligibility (i.e., \emph{scheduling system} agents). 
It is worth noting that, contrary to the other agents, the first two kinds of agents (i.e., assistant agents and physician agents) may not be under complete control of the stakeholders who develop and own the application. 

We introduce a \emph{vaccination centre organisation} to delimit the \emph{vaccination} application and to provide scope for the adherence to regulations and behaviours for both artificial and human agents that are part of this structure.
To this end, the organisation specifies  roles and norms, whereby the roles are used to structure agent responsibilities, and the norms (i.e., duties, rights, and interdictions) regulate the vaccination application. Agents with a given role are expected to fulfil the corresponding norms.
The vaccination centre is, in turn, part of the \emph{health service organisation} of a particular state, in which \emph{clinic organisations} complement the normative framework provided by state and vaccine centres.
In addition to norms, the definition of the organisation may impose hard constraints on its composition that should be enforced by service policies. For example, by 
stating an upper limit on the number of agents that can adopt specific roles, the vaccination application may consequently limit the number of patients or physicians that may enter the organisation.

\subsection{Illustrative Use of the Conceptual Framework with our Motivating Scenario}
The following paragraphs describe the use of the conceptual model in situations derived from the motivating scenario. For narrative convenience, we use the terms \emph{obligation}, \emph{permission}, and \emph{authority} in an informal sense.

\bigskip

\subsubsection{Initialising the Vaccination Application}~\label{subsubsec:init-application}
At the launch of the application, the \emph{vaccination centre organisation} is created, by endowing the agents that support the business processes within the \emph{vaccination centre organisation} with the roles necessary to fulfil their goals. 
The definition of the organisation (e.g., the roles and distribution of norms on the roles) is published as a web resource in a machine readable and understandable format, accessible to any agent wishing to become a member of that organisation. The current state of the organisation (i.e., which agent is assigned the various roles) is published and updated as necessary, over the entire lifetime of the organisation. 
The \emph{freezer agent} adopts the \emph{guard role}, which results in it being assigned the duty of managing access to the inventory of COVID-19 vaccine doses stored in the freezer. It obtains the permission to use the robotic arm to retrieve a vaccine dose when asked, and to deliver it to the staff.
The \emph{manager agent} is assigned the \emph{organiser role}, and consequently inherits the obligation to compile lists of eligible patients based on the patient data and the vaccination eligibility policy. 
The \emph{data agent} adopts the \emph{collector role} and obtains the authority to collect personal information about each patient requesting a vaccination appointment; it also has the obligation to verify the patient's eligibility for receiving the vaccine as well as the obligation to solicit patients through dissemination channels when vaccine doses and scheduling slots are available.

\bigskip

\subsubsection{Joining the Vaccination Centre Organisation}~\label{subsubsec:joining-centre}
When a patient obtains the credentials to access and use the vaccination application, the \emph{assistant agent} acting on behalf of the patient is provided access to the web resource describing the organisation. After reasoning over its obligations and authorities, as imposed by the \emph{vaccination centre organisation}, the agent decides to adopt the role. 
The \emph{assistant agent} subsequently acquires the obligation to provide access to the patient medical data. This role may create internal conflicts between preferences provided by the patient and the obligations assumed when the agent took on the \emph{patient assistant role}.
After accessing and reasoning about the description of the \emph{vaccination centre organisation}, the \emph{physician agent} discovers that it has the obligation to coordinate with agents that are assigned to other roles. To assume the \emph{medical practitioner role}, the \emph{physician agent} must authenticate itself; upon adopting the role, it captures the associated permissions, obligations, and authorisation for further decision making.
The same process of role adoption applies to other agents. 

\bigskip

\subsubsection{Assessing Patient Eligibility}~\label{subsubsec:assessing-eligibility}
While assigned to the \emph{organiser role}, the \emph{manager agent} takes into account its preferences in defining the patient information collection policy, and sends it to the agent with the \emph{collector role} (as stated by the organisation definition). Fulfilling its obligation, the agent checks the eligibility of all arriving patients so that each dose is only administered to an eligible patient and that doses are administered before their expiry date.\footnote{In a practical scenario, we would not expect 100\% compliance with this constraint, but rather that the number of excess or wasted doses (relative to administered doses) does not exceed a specified threshold.}
To fulfil its goals, the agent therefore requests that agents adopting the new \emph{patient assistant role} share the necessary patient personal data. 
It is worth noting that agents with the \emph{collector role} need to consider the obligations stated by the organisation as soft constraints, and identify contexts in which these constraints may be relaxed.
Further complications may arise if any of the agents attempt to negotiate relaxations of these obligations, either in anticipation of, or after a (perceived or factual) violation.
For example, an \emph{assistant agent} can negotiate an exception for a potential obligation violation by using computational models and algorithms of formal argumentation where the \emph{assistant agent} believes that the data it has for its patient satisfies the eligibility criteria. 
On behalf of the organisation, the agent that has adopted the \emph{organiser role} is in charge of the definition of the eligibility policy, and consequently may interact with the agent in charge of the data collection by granting or denying the request for an exception, or even by updating the organisation's norms in order to accept the request.

\bigskip

\subsubsection{Administering the Vaccine}~\label{subsubsec:administer-vaccine}
When administering the vaccine, the agent with the \emph{physician role} must respect the priority order for vaccine administration as defined by the agent in charge of the collection; for example, the elderly and vulnerable population must be vaccinated first, unless respecting the priority order implies wasting the dose. Importantly, the physician must not violate this priority order by, for example, preferentially vaccinating friends or relatives.
In some cases, the \emph{physician agent} must choose between either administering a vaccine dose despite the eligibility status being uncertain, or allowing the dose to expire.
For example, existing regulations indicate that administering a vaccine dose to close relatives is impermissible, but the agent may conclude that in accordance with practitioner norms (such as the Hippocratic Oath), it is preferable if vaccines are not wasted.
In such cases, the agent may need to prepare a defence strategy to avoid sanctioning, for example, via argumentation approaches~\cite{baroni2018handbook,Bench-Capon2009}.
These issues merely relate to the permissions needed for the obligations implied by opening the fridge. Additional challenges arise when considering the complete socio-technical system, including electronic health record access \citep{10.1145/3365664} and supply chain integration \citep{5331908}.


\section{Challenges and Opportunities}
\label{challenges-and-opportunities}

\begin{figure}[t]
    \centering
    \includegraphics[width=0.4\linewidth]{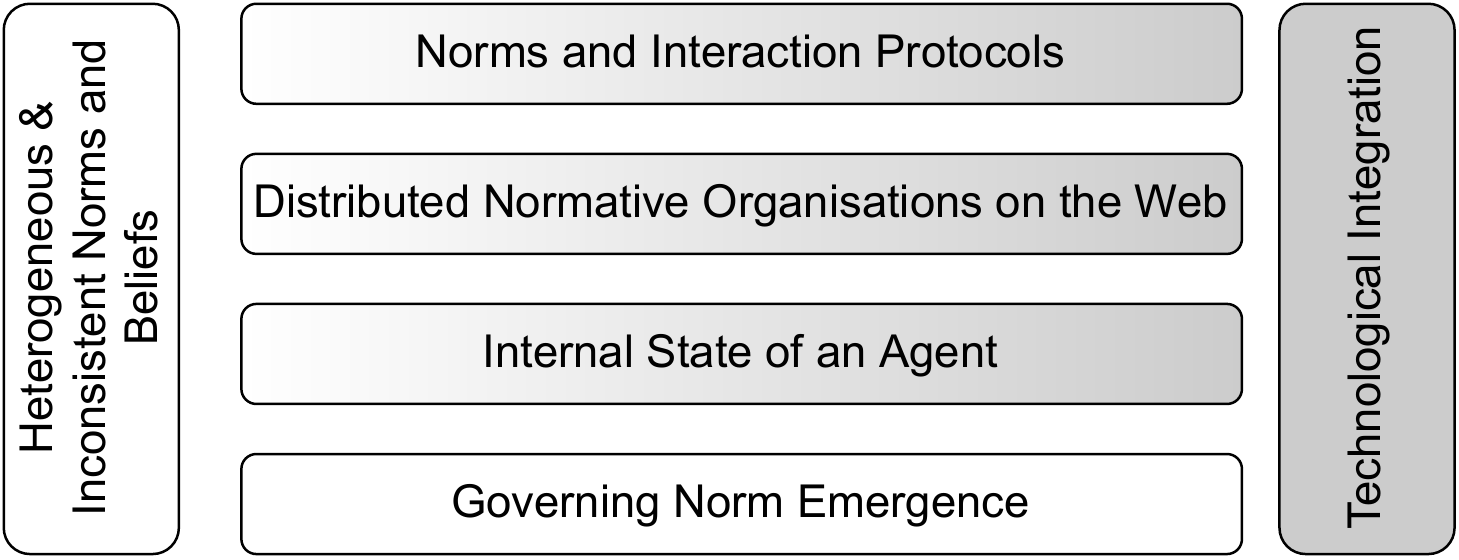}
    \caption{Overview: Research challenges for normative MAS and the Web} 
	\label{fig:challenges}
\end{figure}

We now present research challenges (i.e., technical limitations of existing proposals) and opportunities (i.e., open research questions) related to normative agents on the Web. The four horizontal challenges that characterise the contributions of norm-based multiagent systems for the Web are described below, in addition to two orthogonal challenges that need to be tackled in order to address the horizontal issues (and illustrated in Figure~\ref{fig:challenges}).
For each area, we label the challenges in the context of limitations of the state of the art and subsequently identify future research opportunities. 
Figure~\ref{fig:challenges} also indicates the practical maturity of each challenge, from \emph{nascent} (blue sky challenges with basic research potential, using a white background) via \emph{developing} (basic research with immediate practical potential, using a white-grey gradient background) to \emph{practical} (challenges that can be addressed primarily from an engineering perspective using a grey background).

\subsection{Relating Norms and Interaction Protocols}
\subsubsection*{Challenges}
A broad challenge in engineering normative MAS is that we need a way to operationalise norms in the sense of giving them a computational interpretation. Interaction protocols characterise interactions based on message order and occurrence~-- that is, in operational terms. However, it is nontrivial to produce protocols that are as flexible as necessary, yet enactable in a decentralised manner, while at the same time being verifiably correct. Although the W3C provides Web-based standards for retrieving and querying machine-readable data, these standards do not cater for usage constraints, such as access policies, intellectual property rights, and privacy preferences.
In our scenario, an agent may, for example, want to decide with whom and in which context it shares its vaccination status.
Existing work on interaction protocols~\citep{FIPA-IP} largely focuses on request-response interactions and imposes restrictions on computation for scenarios involving the interaction of three or more parties \citep{Ferrando+19:enactability,JAIR-20:Langeval}. In particular, traditional approaches entwine control flow details into the protocol, thereby making it difficult to separate them from the content, for which a declarative meaning can be specified. Prior work on specifying protocols based on norms (commitments) \citep{Chopra+Singh-04,JAAMAS-Algebra-07} was hindered by the lack of declarative specification of the constraints on messages.
More recent approaches describe causality and integrity constraints on messages declaratively~\citep{AAMAS-BSPL-11}; such protocols (whilst sufficiently flexible to support all enactments of the stated norms) can grow quite large~\citep{AAAI-20:Clouseau}, but emerging verification approaches aim to tackle this challenge~\cite{IJCAI-21:Tango}.

\subsubsection*{Opportunities}
When it comes to operationalising norms, from a service-provisioning perspective, there is a need to develop policy-aware querying and data retrieval protocols; whereas from an agent interaction perspective, norms should be mapped both to the agent platform and the environment. This raises several important questions, including:
\begin{inparaenum}[(i)]
\item How can we design norm-aware dynamic interaction protocols?
\item How can existing querying and data retrieval protocols be extended, such that they are policy aware?
\item What new languages are needed to facilitate norm governance?
\item How do we model and reason with respect to norm changes and temporal validity?
\end{inparaenum}

When operating in deployed systems modelled according to the conceptual framework described in Section~\ref{sec:conceptual-framework}, these interaction protocols also need to take into account that the autonomous agent layer includes  human agents which might have conflicting requirements, and therefore strategies might need to be employed in order to resolve conflicts. This raises a number of additional questions that need to be addressed, including:
\begin{inparaenum}[(i)]
\item How do we ensure protocol compliance by human agents?
\item How do we model protocols that implement persuasion?
\item What mechanisms do we use to resolve conflicting requirements?
\end{inparaenum}

\subsection{Distributed Normative MAS and Open Organisations on the Web}
\subsubsection*{Challenges}
Organisations, institutions, and contracts are useful abstractions to structure norms and make them accessible to agents. Although agents have the choice of joining such structures, they may be subject to conditions that regulate their admission (and exit), as well as there being an expectation to comply with the organisation's norms.
Due to the scale of the Web, numerous permanent, ephemeral, or evolving structures may exist. Consequently, an agent needs to be able to discover and reason about such organisations and the corresponding norms.
In the vaccination scenario, for example, an agent may need to be able to discover organisations that model the healthcare systems of other jurisdictions that potential patients may need to refer to, when claiming that they are eligible to receive a specific type of vaccine dose.
Ontologies facilitate the discovery of services \citep{paolucci:02}, and their use as a means to represent organisations is promising.
A major challenge related to the distributed management of such structures~\citep{boissier2020multi} is to monitor and enforce norm compliance, and to instantiate organisations, agents, or complete multiagent (sub-)systems at run-time on the Web, which is an emerging line of research in the MAS community~\cite{amaral2020towards,10.1007/978-3-030-51417-4_13,aldewereld-et-al:2010}.
Another challenge is that agents require abstractions and mechanisms to build and adapt organisations on the fly~\citep{WWW-16:IOSE}. Additionally, an agent may participate in multiple structures that operate at different timescales and scopes, and hence accommodating their diversity is nontrivial.

\subsubsection*{Opportunities}
Addressing the above challenges requires answers to the following research questions, in the context of Web-based and WoT-based technology ecosystems:
\begin{inparaenum}[(i)]
\item How can agents discover organisational norms on the Web?
\item How can norm compliance be monitored and enforced in dynamic scenarios, in which agents, organisations, or entire (sub-)MAS are instantiated at run-time?
\item How can autonomous agents create and change organisations \emph{on-the-fly}?
\item How can normative organisations accommodate agents that participate in multiple organisations, with potentially inconsistent norms, and partial semantic interoperability?
\item How can compliance checking and enforcement approaches that are prevalent in the information systems literature be adapted and applied to normative MAS on the Web?
\end{inparaenum}

\subsection{Internal State of an Agent and Norms}
\subsubsection*{Challenges}
An agent should be able reason about norms, taking into consideration its internal state (e.g., its beliefs, goals, and intentions), and explain its normative reasoning to others. This is, for example, important for the \emph{administering the vaccine} scenario (Section~\ref{subsubsec:administer-vaccine}), when a decision is to be made about whether to administer a vaccine to a patient whose eligibility status is uncertain, using for example qualitative methods~\citep{parsons-qualitative}, argumentation~\citep{baroni2018handbook} or sub-symbolic techniques such as classification or Bayesian networks.
Several languages express and support automated reasoning about agent internals, such as beliefs, desires or goals, and intentions. However, challenges exist when it comes to reasoning not only with respect to goals, but privacy preferences, regulatory constraints, and norms. The problems here not only relate to knowledge representation~-- how to represent all these aspects so they are accessible to an agent~-- but also to the impact such enrichment may have on deliberation performance. The most popular (symbolic) agent architecture is practical reasoning by means of beliefs, desires and intention (BDI), but this offers no guarantees about how long the deliberation cycle may take. Indeed this falls to the developer, in as much as they can endeavour to keep rules relatively simple whilst limiting the number of overlapping conditions, but the actual response time is out of their hands.
Additionally, there is the more significant problem that arises due to the fact that agent architecture research has primarily focused on agent internals. Irrespective of whether they are comprised of symbolic or sub-symbolic aspects (such as reasoning, reinforcement or probabilistic learning), such architectures are not normally conceived or designed for interfacing with non-agent technological frameworks and their underlying abstractions (for example as provided by the W3C Web architecture and related standards).

\subsubsection*{Opportunities}
From a norms perspective, important open questions in specification and enforcement include:
\begin{inparaenum}[(i)]
\item How can we ensure consistent representation of and adherence to norms?
\item How should a governance architecture be designed in which rational agents are incentivised to comply with norms?
\item Can a norm violation be excused, based on explanation or argumentation? 
\item Could transparency facilitate the persuasiveness of the explanation or argument? 
\item How do we ensure that an agent is aware of the implications of violating a norm? 
\item How do we cater for agents that are not rational (as such agents may not be designed for criteria such as rationality)?
\item How should performance limitations affect normative decision-making in practical reasoning architectures?
\item How can normative reasoning be implemented in real-world agent architectures (which may only be agent-oriented in the broader sense)?
\item How can agent architectures be better integrated with web technologies and standards?
\end{inparaenum}

\subsection{Governing Norm Emergence}
\subsubsection*{Challenges}
Approaches for the governance of norm emergence are dependent on the capabilities of the agents in a MAS, bearing in mind that population properties may not be homogeneous.
In our example scenario, the governance of norm emergence is, for example, important to facilitate vaccinations (i.e., the belief that getting vaccinated is, while typically not mandatory, good for one's health and more broadly for the public health at large), and to balance ``hard'' rules and ``soft'' recommendations to decrease the spread of COVID-19. The challenges here include modelling and managing the spread of beliefs and counter-beliefs, the potential resolution of contrary positions through argumentation, and how to make hard and soft policies accessible to different agent architectures with different reasoning capabilities.
We differentiate between a decentralised approach to norm emergence with implicit norms, where the norms emerge through the interactions of agents~-- \citep{IJCAI-18:Poros} is one example of such a scheme~-- and various centralised approaches to the governance of norm emergence~\citep{normEmergence}, which latter we classify by adapting the oversight terminology put forward by the~\citet[\S{}B.II.1.1]{aihleg}:
\begin{inparaenum}[(i)]
\item an external agency observes the behaviour of the population to identify patterns of behaviour and revise norms imposed by that agency to optimise for system goals (external agent/human on-the-loop) -- for example~\citet{DBLP:journals/taas/MoralesLRVW15} look at individual norms in isolation -- while the general case of revising a consistent body of norms remains open;
\item agents propose norm revisions to an external agency, which then implements them subject to an assessment of how those revisions contribute towards system goals (external agent/human in-command), which also remains open; and
\item agents propose norm revisions and system participants, which may include humans, use an internal decision-making mechanism to establish which changes will be implemented (internal agents/humans in-the-loop). The uHelp system illustrates some preliminary steps in this direction~\citep{DBLP:journals/sncs/OsmanCSSG21}, but relegates software agents to a supporting role.
\end{inparaenum}
The human-in-the loop, human-in-command, human-on-the-loop approaches are closely related to the different strategies that regulate persuasive dialogues~\cite{Boella2004e}, where commands introduce new obligations, whilst suggestions introduce new beliefs. However, open challenges relate to devising persuasion strategies and the corresponding obligations.

\subsubsection*{Opportunities}
In order to govern norm emergence in Web-based MAS and their socio-technical contexts, one needs to answer~-- for example~-- the following questions:

\begin{inparaenum}[(i)]
    \item How can the emergence of norms in conjunction with governance decisions be monitored and managed in Web-based MAS?
    \item What roles do human-on-the-loop, human-in-the-loop, and human-in-command approaches play in the context of the preceding question, and what are the engineering implications of these different human interaction approaches?
    \item What is necessary to maintain alignment of the (evolving) value preferences of participants with the norms that govern them: when does one norm change become many changes?
    \item Which collective decision-making mechanisms are best suited for all agent and for mixed human-agent systems?
    \item What is the appropriate capacity for agent reasoning about (self-)governance? Is wanting to do something that is prohibited, or not wanting to do something that is obliged a sufficient statement of intention?
    \item How can an agent and/or a human evaluate the consequences of norm revisions? Will they create a fresh problem while resolving the current one?
    \item How can oscillatory norm change be prevented? An agent-dominated system could potentially change faster than a human-in-command can evaluate the changes.
\end{inparaenum}

\subsection{Heterogeneous and Inconsistent Norms and Beliefs}
\subsubsection*{Challenges}
In heterogeneous information systems, we cannot reasonably assume that norms and policies are globally accepted and thus agents 
may hold inconsistent beliefs about them.
For example, in our vaccination scenario, vaccine administration policies, eligibility requirements, and IT system landscapes may differ between two federated states A and B.  However, a patient who moves permanently from A to B should ideally be able to receive the first vaccine dose in state A, and a dose of a matching or complementary type in state B after a reasonable time interval.
For aligning norms and policies of sub-entities, reaching (partial) agreements in the face of conflicting beliefs regarding norms and policies is an important challenge that needs to be tackled to enable normative distributed MAS on the Web; using, for example, long-running lines of research on agreement technologies~\citep{10.5555/2431387} and formal argumentation~\citep{baroni2018handbook}.
Currently, the body of research on belief revision and argumentation-based reasoning is, however, poorly integrated with practical engineering perspectives; standardisation efforts like the specification of an \emph{argument interchange format}~\cite{Rahwan2009} exist, but have not found substantial adoption.

\subsubsection*{Opportunities}
There are several open questions when it comes to enabling reasoning and decision-making in the face of inconsistent norms and beliefs of agents on the Web, including:
\begin{inparaenum}[(i)]
    \item To what extent is there a practical need for engineering abstractions that treat conflict and inconsistency in the context of a normative Web (of Things)?
    \item What systematic approaches to drawing inferences and making decisions in a Web (of Things) governance context can be designed, implemented, and standardised as software engineering abstractions?
    \item How can existing research on belief revision and argumentation-based reasoning be made more accessible both from an engineering and a standardisation perspective?
\end{inparaenum}

\subsection{Technological Integration}
\subsubsection*{Challenges}
In order to facilitate the practical applicability of research on norms and policies for autonomous agents on the Web, it is crucial to build bridges across the technology ecosystems of the different communities.
Section~\ref{background} provides an overview of the technology ecosystems that have emerged from the MAS, Semantic Web, and WoT communities. To summarise, in the WoT community, engineering-oriented work has been conducted in a highly practice-oriented manner, in close alignment with industry practitioners as well as standardisation bodies such as the W3C.
An example of practice-oriented work can be observed though W3C IoT standards that feature an abstract architecture~\citep{W3C_WoT_Arch:20} and an interface specification (W3C WoT Scripting API)~\citep{wot-scripting-api:20}, supported by a JavaScript reference implementation.\footnote{\url{https://github.com/eclipse/thingweb.node-wot}}
Research on engineering autonomous agents and MAS has primarily gained traction within the academic community~\citep{engineering-gsi-article-2019}, and standardisation attempts such as  FIPA\footnote{\url{http://www.fipa.org/}.} have lacked significant adoption. Adjusting agent-oriented programming and software engineering approaches to better serve the Semantic Web and WoT communities is a way for the MAS community to move their engineering research closer to practice.
This lets us conclude that while each of the communities has its own thriving technology ecosystem, a key challenge lies in integrating these ecosystems, which exhibit different degrees of practical maturity.

\subsubsection*{Opportunities}
The above observations raise two questions:
\begin{inparaenum}[(i)]
    \item How can the technology ecosystems of (normative) MAS and the Semantic Web be integrated with the WoT, and in particular with the W3C Web architecture?
    \item How can issues of practical maturity be mitigated (by the integration strategy)?
\end{inparaenum}
With respect to \emph{(i)}, we argue that an integration strategy can employ a combination of two approaches across two dimensions that requires pragmatic trade-offs, considering the discrepancies between the technology ecosystems and their underlying conceptual abstractions.

\noindent \textbf{Approach 1: Full-Fledged Framework Adoption.} In order to facilitate implementations that build on research in the different communities, interfaces that integrate Semantic Web, MAOP, and WoT technologies can be devised that either re-implement their abstractions or integrate technology frameworks and specification languages~\cite{10.1007/978-3-319-91899-0_8,DBLP:conf/emas/CiorteaBR18,ciortea:hal-02196903}. A benefit of this approach is that it facilitates the adoption of powerful abstractions and technology ecosystems developed in these communities. A disadvantage, however, is that this approach can cause a high technological overhead. \\
\noindent \textbf{Approach 2: Modular Abstraction Adoption.} In order to facilitate implementations that build on technology stacks established in industry, \emph{minimally viable abstractions} on norms and autonomous agents can be implemented as reusable modules in mainstream programming languages; or alternatively, specific features of complex technology platforms can be exposed as service-oriented interfaces. This strategy resembles a call to action as made in Logan's \emph{Agent Programming Manifesto}~\cite{logan2018agent} and allows for deliberate trade-offs between conceptual richness and practical feasibility by avoiding the overhead (on conventional developers) of having to learn unfamiliar programming paradigms. For example, one may adopt JaCaMo's capabilities for modelling organisations and artefacts via a Java-based technology stack, and defer adopting Jason \cite{Bordini-et-al07:book} since it involves a custom language for agent-oriented programming.

Broadly, the two approaches can be considered analogous to the \emph{integrated system} (Approach 1) and \emph{best of breed} (Approach 2) strategies for implementing large scale enterprise systems~\cite{bestbreed}. In an actual implementation scenario, these approaches represent the extremes of a scale with valuable trade-offs in between.
We suggest that this trade-off is initially made from a conceptual perspective (\emph{Which programming abstractions are useful in a given scenario?}) and followed using a technology perspective (\emph{Which technologies do I want to use to implement these abstractions?}).

With respect to the second question, we argue that the strategy should prioritise mature technologies, and if necessary re-implement the requisite abstractions in technology stacks that are established in industry practice.
Specifically, we might consider the WoT standards and technologies as a mature foundation on which to place Semantic Web and MAS technologies.
An example of a synergy is in extending WoT servients to autonomous agents without necessarily committing to a BDI architecture for those agents.
\vspace{-5pt}

\section{Conclusion}\label{conclusion}
This paper discusses the relevance of norms, policies, and preferences for governing complex socio-technical multiagent systems on the Web. The key challenge~-- the conceptual and technological integration of normative concepts with WoT abstractions and systematic evaluation of the practical usefulness of the integration results~-- is aligned with the general challenge for autonomous agents on the Web to transfer the rich theoretical achievements of the broader MAS community to the practical and engineering-oriented WoT community, and to facilitate real-world applications at scale.
While the challenge of transferring research on normative agents and multiagent systems into engineering practice is well-known and generally acknowledged, this paper has taken the emergence of new Web standards, as well as the increased research interest in Web-based MAS, as a starting point to provide a new and broad perspective on it, with a focus on the Web and Web of Things Architecture standards.
In this context, the paper proposes a conceptual framework that serves to define the role played by various norms, policies and preferences when it comes to complex socio-technical MAS on the Web, and demonstrated it via a simple but realistic scenario.

In addition, the paper provides a research roadmap outlining the technical and theoretical research challenges and opportunities to support complex socio-technical MAS governance on the Web. In particular, this roadmap calls for:
\begin{inparaenum}[(i)]
\item relating norms and interaction protocols; 
\item incorporating normative organisations and norm governance approaches into WoT architectures and standards;
\item combining agent reasoning to relate policies, preferences, and norms;
\item tackling the emergence of norms for flexible governance;
\item designing reasoning methods about norms in the face of inconsistency; and
\item cautiously advancing Semantic Web and (normative) MAS tools and frameworks into practice via the WoT.
\end{inparaenum}

\subsubsection*{Acknowledgments}
%
We thank the anonymous reviewers.
Timotheus Kampik is supported by the Wallenberg AI, Autonomous Systems and Software Program (WASP) funded by the Knut and Alice Wallenberg Foundation. 
Adnane Mansour, Olivier Boissier, and Antoine Zimmermann are supported by the French National Research Agency under grant ANR-19-CE23-00030 (project HyperAgents).
Sabrina Kirrane is funded by the FWF Austrian Science Fund and the Internet Foundation Austria under the FWF Elise Richter and netidee SCIENCE programmes as project number V 759-N. 
Munindar P.~Singh acknowledges support from the US National Science Foundation under grant IIS-1908374.

\bibliographystyle{ACM-Reference-Format}
\bibliography{norms+policies}
\end{document}